# Decays of the $f_0(1370)$ scalar glueball candidate in pp Central Exclusive Production (CEP) and in antiproton annihilations at rest

*Ugo Gastaldi*

INFN-Ferrara, Via Saragat 1, 44122 Ferrara, Italy

**Abstract.** Decays into $\pi^+\pi^-$ of the $f_0(1370)$ are the main source of an isolated structure localized between 1.2 and 1.5 GeV in the $\pi^+\pi^-$ mass spectrum measured in pp Central Exclusive Production (CEP) at $\sqrt{s}$=200 GeV at very low four momentum transfer ltl by the STAR experiment. These data confirm in the $\pi^+\pi^-$ decay channel the existence of the $f_0(1370)$ as an isolated well identified structure previously observed in $K^+K^-$, $K_sK_s$, $\pi^+\pi^-\pi^+\pi^-$, $2\pi^0\pi^+\pi^-$ and $4\pi^0$ decays measured in pbar annihilations at rest. The ensemble of these data point at a high gg content of the $f_0(1370)$. CEP interactions at higher energies favour production of $0^{++}$ and $2^{++}$ mesons. Selection of events with lower ltl at both proton vertices suppresses $2^{++}$ structures. LHC runs dedicated to pp CEP measurements at low ltl could then provide a unique source of all the low energy scalars. This would make it clear if and where scalar gluonium is resident and the nature (composition in terms of qqbar, qqqbarqbar, qqbar-qqbar and gg) of $f_0(500)$, $f_0(980)$, $f_0(1370)$, $f_0(1500)$ and $f_0(1710)$.

## 1 Introduction

The possibility of the existence of glueballs is a basic qualitative prediction of quantum chromodynamics (QCD)[1,2]. The lowest lying glueball states are expected to have the same $0^{++}$ $J^{PC}$ quantum numbers as the vacuum. While for the $0^{-+}$ pseudoscalar, $1^{--}$ vector and $2^{++}$ tensor meson nonets there are two observed isoscalar partners for two places in each nonet (respectively η and η', ω and φ, $f_2(1270)$ and $f'_2(1525)$, there are 4 or 5 observed isoscalar mesons (respectively σ, $f_0(980)$, $f_0(1370)$, $f_0(1500)$ and $f_0(1710)$ for the two places of the isoscalar members of the $0^{++}$ ground state nonet. The candidates for the two places are 4 or 5 depending whether σ and $f_0(1370)$ are considered as distinct separated objects or they are part of a single continuum. The existence of the σ meson is considered established since some years (see [3] for a review) and σ is currently called $f_0(500)$. Doubts instead concern the existence of the $f_0(1370)$ as an individual isolated structure (see [4] for a review). Concerning the nature of the σ and $f_0(980)$ $0^{++}$ mesons, various scenarios envisaged in the literature include (because of mixing) the possibility of a qqbar, qqqbarqbar, qqbar-qqbar and gg content in their wave functions[3-7]. If the $0^{++}$ isoscalars are 5 it is more difficult to exclude the hypothesis of the presence (or even dominance) of a gg content in some of them. Production in glue rich processes (central production mediated by double pomeron exchange, pbar annihilations, J/ψ and ψ' radiative decays, heavy meson decays) and absence in γγ production and meson exchange are criteria useful to characterize the gg content of the scalars [8]. Relative decay branching ratios into σσ, ρρ, ππ, KKbar, ηη for the three heavier scalars, and also into ηη' for $f_0(1500)$ and $f_0(1710)$, give other selection criteria to identify their gg content [8-11]. Currently there is no consensus concerning the experimental observation of a scalar glueball nor on the possible gg content in the scalars experimentally observed (for reviews see [4,12-17]).

An isolated $\pi^+\pi^-$ peak practically without background and well separated from the $f_0(980)$ signal by a valley nearly devoid of events present in pp CEP data of the STAR experiment [18] has been interpreted as a clear manifestation of the $f_0(1370)$ meson as an isolated structure [19,20]. This interpretation is based on STAR data [18, 21-25], on the scaling laws of CEP [8,26] and on results of the AFS experiment [27,28] at the CERN ISR. This points are discussed in section 2.

In section 3 we recall previous data where the $f_0(1370)$ is present as an isolated clearly visible structure in $K^+K^-$, $K_sK_s$, $\pi^+\pi^-\pi^+\pi^-$, $2\pi^0\pi^+\pi^-$ and $4\pi^0$ decays measured in pbar annihilations at rest. Some of these data have been ignored in the PDG compilations of the last 20 years concerning $f_0(1370)$ and $f_0(1500)$, some are used in data averages to give mass, width and decay branching ratios of $f_0(1500)$, and are discarded in the $f_0(1370)$ section. Besides establishing –together with the STAR data- the existence of the $f_0(1370)$ as an isolated structure, these data are essential to extract the production rates and decay branching ratios of both $f_0(1370)$ and $f_0(1500)$, in order to identify their gg content. The orders of magnitude of the decay branching ratios of $f_0(1370)$ into $\sigma\sigma$, $\rho\rho$, $\pi\pi$, $K\bar{K}$, $\eta\eta$ relative to $\pi\pi$ decays are 6, 3, 1, 1, 0.02. These data point at a large gg content of the $f_0(1370)$ meson and need a definitive confirmation by CEP measurements at LHC, also because there are conflicting results (based on low energy CEP data) of the WA102 experiment.

Section 4 gives conclusions, stresses the importance of the assessment of the properties of the $f_0(1370)$ for the observation of a scalar glueball and highlights the substantial improvements achievable by pp CEP experiments at LHC in the medium and longer term for the spectroscopy of all low energy scalars.

## 2  $f_0(1370)$ from $\pi^+\pi^-$ decays in pp Central Exclusive Production

Fig.1 [18] shows row data of $\pi^+\pi^-$ pairs in pp CEP data at $\sqrt{s}$=200 GeV produced by the STAR experiment at RICH in a 2009 run with $\beta^*$=20 m setting of the storage rings and precision forward detectors inserted in two sets of roman pots positioned at 55.5 and 58.5 m distance from the central detector at its two sides . This configuration of machine and forward detectors gives access to the kinematical region for the four momentum transfer square l$t_1$l and l$t_2$l at both proton vertices with 0.003 < l$t_1$l , l$t_2$l  < 0.03 GeV$^2$. This l$t_1$l, l$t_2$l acceptance window of the 2009 STAR run -shown in fig. 2- is the ltl window with lowest values so far exploited in pp CEP experiments.

Fig. 3 [23-25] shows row data of $\pi^+\pi^-$ pairs in pp CEP data at $\sqrt{s}$=200 GeV produced by the STAR experiment at RICH in a 2015 run with $\beta^*$=0.85 m setting of the storage rings and the two sets of roman pots positioned at the two sides of the central detector at 15.8 and 17.6 m distance.  This second configuration of the machine and forward detectors gives access to the kinematical region with four momentum transfer square l$t_1$l and l$t_2$l at both proton vertices in the window  0.03 < l$t_1$l , l$t_2$l  < 0.3 GeV$^2$ (also shown in fig.2). The sharp drop in fig. 3 at 1 GeV is due to the $f_0(980)$ scalar meson; the strong peak near 1.3 GeV is due to the $f_2(1270)$ tensor meson. These two signals permit to check the energy calibration of the experiment.  The sharp drop near 1 GeV is also present in fig.1, where the energy region comprised between 1 and 1.2 GeV is nearly empty of events, while a peak with about 60 events is present in the energy window 1.2-1.5 GeV. The peak in fig.1 between 1.2 and

1.5 GeV has been interpreted in [19,20] as due essentially to the $f_0(1370)$.

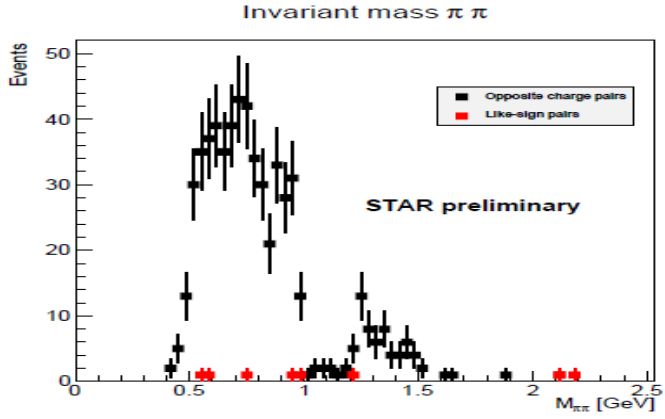

**Fig. 1.** $\pi^+\pi^-$ mass plot of STAR pp CEP raw data of the 200 GeV 2009 run with 0.003<|$t_1$|, |$t_2$|<0.03 GeV$^2$ kinematic coverage of both scattered protons (Fig. from [18]).

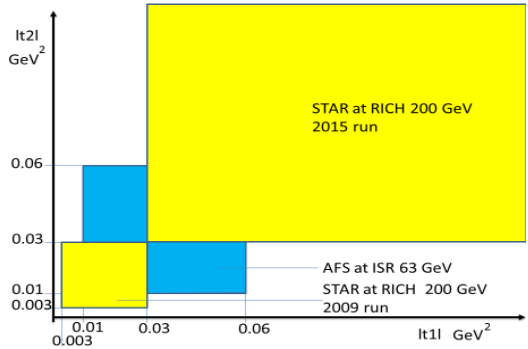

**Fig. 2.** Kinematical acceptances in the |$t_1$|, |$t_2$| plane of the AFS [27,28] and STAR [18,23] pp CEP experiments

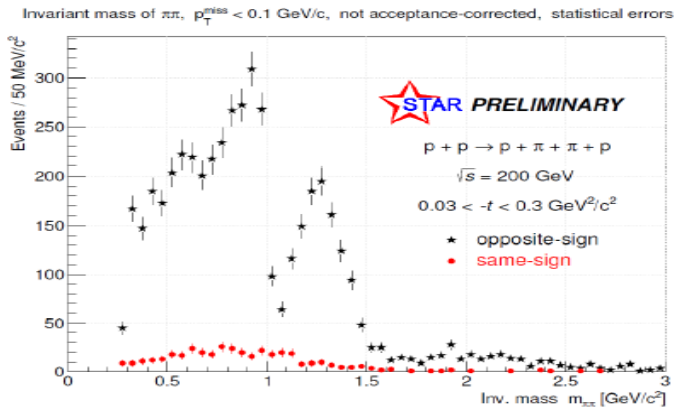

**Fig. 3.** $\pi^+\pi^-$ mass plot of STAR pp CEP raw data of the 200 GeV 2012 run with 0.03 <|$t_1$|, |$t_2$|<0.3 GeV$^2$ kinematic coverage of both scattered protons (Fig. from [23]).

The angular distribution of the $\pi^+\pi^-$ events in the peak is not available, but the S-wave nature of the signal can be deduced by the fact that already at $\sqrt{s}= 63$ GeV the $\pi^+\pi^-$ signal of the AFS experiment was dominantly S-wave up to 1.7 GeV [27,28,16] with a higher ltl window than STAR and because, as shown by comparing the 2009 and 2012 STAR data, lowering the ltl window suppresses dramatically the $2^{++}$ $f_2(1270)$ signal. The structure in fig.1 between 1.2 and 1.5 GeV centered around 1.35 GeV is too large to be associated to the $f_0(1500)$, which has a width of the order of 100 MeV, and is expected at 1.5 GeV. The statistics is too low to extract meaningfully a ratio between the $f_0(1370)$ and the $f_0(1500)$ contributions. Since most of the events in the peak occur at masses below 1.5 GeV, the $f_0(1500)$ may contribute to the spectrum via a destructive interference with the $f_0(1370)$ amplitude, as suggested in the analysis of the AFS data [28]. Independently of the interpretation of the structure in the 1.2-1.5 GeV energy region, quite noticeable is the fact that the STAR $\pi^+\pi^-$ spectrum of fig.1 drops nearly to zero at 1 GeV. This may be the result of the interference of the amplitudes of the low energy tail of the $f_0(1370)$ with the high energy part of the $f_0(980)$ plus the effect of the $K\bar{K}$ threshold, but very likely it might be due to the vanishing of the S.wave continuum for events selected in the low ltl kinematical region. The S-wave continuum, which is usually invoked with its destructive interference with the $f_0(980)$ amplitude to generate the drop at 1 GeV, is drastically reduced in comparison to the AFS data. It looks like the $\sigma$ meson, which generates the broad structure above 0.5 GeV, is confined below 1 GeV. In other words, under this hypothesis the "red dragon" glueball proposed by Minkowski and Ochs [29], which features a low energy body centered at about 0.6 GeV and a head extending below the $f_0(1500)$, would be split into two separate parts, the $\sigma$ and the relatively narrow $f_0(1370)$.

## 3  $f_0(1370)$ from $K\bar{K}$ and 4 pion decays in $\bar{p}$ annihilations at rest

Decays of $f_0(1370)$ into $K^+K^-$ and into $K_sK_s$ are directly observable in $\pi^0K^+K^-$, $\pi^-K_sK_s$ and $\pi^0K_sK_s$ Dalitz plots and in the respective $K^+K^-$ and $K_sK_s$ mass plots of $p\bar{p}$ annihilations at rest in liquid $H_2$ and $n\bar{p}$ annihilations at rest in liquid $D_2$ targets (see fig. 4 from ref. [30]). $K^+K^-$ decays of $f_0(1370)$ are the main source of the diagonal $K^+K^-$ band comprised between the $K^{*+}$ and $K^{*-}$ bands in the $\pi^0K^+K^-$ Dalitz plot of the top left plate of fig.4 and of the peak at 1.4 GeV in the $K^+K^-$ mass plot of the top right plate of fig. 4. Decays into $K^+K^-$ of the $f_0(1500)$ are not easily visible in ppbar annihilations in liquid $H_2$ targets [31-33] (neither in the $\pi^0K^+K^-$ Dalitz plot, nor in the $K^+K^-$ mass plot) because the $f_0(1500)$ signal is weak in liquid $H_2$ targets, since it is only partially produced from S-wave initial atomic states (while the $f_0(1370)$ signal is dominantly produced from $0^{-+}$ S-wave initial atomic states), and S-wave annihilations dominate in liquid $H_2$ targets (while P-wave annihilations dominate in low density $H_2$ targets) [32-34]. Moreover the weak $f_0(1500)$ signal may be masked by the nearly overlapping $f_2'(1525)$ signal.

Destructive interference of the $f_0(1500)$ amplitude with the $f_0(1370)$ amplitude generates the deep narrow flat valley at $M(K_sK_s)=1.5$ GeV in the $\pi^0K_sK_s$ Dalitz plot and in the $K_sK_s$ mass plot obtained [30] using CERN [35,36] and BNL [37] bubble chamber data (see the bottom left and right plates in fig. 5). This $K_sK_s$ valley in the $\pi^0K_sK_s$ bubble chamber Dalitz plot and in the $K_sK_s$ mass plot represents probably the best evidence of $K\bar{K}$ decays of $f_0(1500)$. The same effect is visible in the high statistics $\pi^0K_lK_l$ Dalitz plot and in the associated $K_lK_l$ mass plot produced by Crystal Barrel [38] (see figs.1 and 2 in [38]), where the valley is half as deep because of background and less mass resolution (notice that ref.[38] uses the frequency of ppbar $\to \pi^0K_sK_s$ annihilations in liquid $H_2$ as frequency of the their ppbar $\to \pi^0K_lK_l$ annihilations).

In the $\pi^-K_sK_s$ Dalitz plot produced [30] using CERN [39] and BNL [40] bubble chamber data (see the central left plate in fig. 4) the effect of the interference of the $f_0(1370)$ band with the $K^{*-}$ bands is clearly noticeable. The $K_sK_s$ decays of $f_0(1370)$ are the main source of the peak at about 1.4 GeV in the $K_sK_s$ mass plot of the central right plate of fig.4. These data represent probably the best direct evidence of $K\bar{K}\ f_0(1370)$ decays. A direct quantitative comparison of $\pi^-K_sK_s$ and $\pi^-2\pi^0$ annihilation data in liquid $D_2$ would help providing relative decay branching ratios of $K\bar{K}$ and $\pi\pi$ decays of $f_0(1370)$, but it is not available.

Data of $p\bar{p}$ annihilations at rest into three final states ($\pi^+\pi^-\pi^0$, $K^+K^-\pi^0$ and $K_sK^+\pi^-/K_sK^-\pi^+$) at 3 densities of the $H_2$ target (liquid $H_2$, NTP gaseous $H_2$ and 5 mbar NT gaseous $H_2$) have been collected and analyzed by the Obelix experiment at LEAR in a coupled channel analysis [32,33]. The spectra at the 3 target densities have markedly different shapes because the relative production of intermediate resonances ($\pi\pi$, $K\bar{K}$, $\pi K$) depends dramatically on the $J^{PC}$ initial atomic states of annihilation, and the $J^{PC}$ fractions of annihilation depend substantially on the target density [34]. The $K^+K^-\pi^0$ data in liquid $H_2$ constrain the width of the $f_0(1370)$ so that the hypothesis of a broad $f_0(1370)$ extending below 1 GeV is naturally discarded, and the $\pi^+\pi^-\pi^0$ data are fit with sensible priors for the $f_0(1370)$ and the $f_0(1500)$ width and mass. The data show that the $f_0(1370)$ is essentially produced only from $J^{PC} = 0^{-+}$ S-wave initial states and that for $f_0(1500)$ there is a relevant production from $J^{PC} = 1^{++}$ P-wave initial states (this feature may explain why $f_0(1500)$ and not $f_0(1370)$ are observed in $p\bar{p}$ annihilations in flight at 900 and at 1640 MeV/c [41]). The coupled channel analysis shows that the ratio between the $K\bar{K}$ and $\pi\pi$ couplings of $f_0(1370)$ is of the order of 1 [32,33].

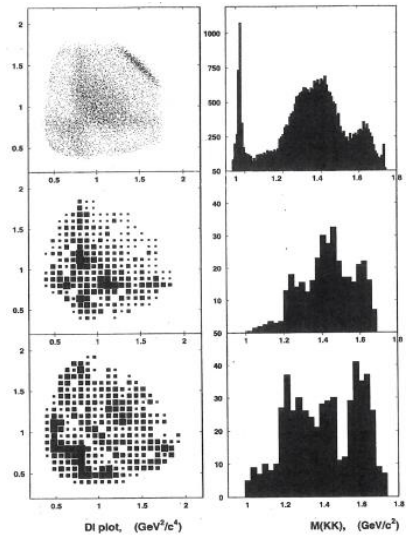

**Fig. 4** $\pi^0K^+K^-$, $\pi^-K_sK_s$, $\pi^0K_sK_s$ Dalitz plots (top, middle and bottom left plates) and respective $K^+K^-$ and $K_sK_s$ mass plots (right plates) of pbar annihilations at rest in liquid $H_2$ and $D_2$ targets (picture derived from figs. 1, 2 and 5 of ref. [30]).

Decays of $f_0(1370)$ into $\pi^+\pi^-\pi^+\pi^-$ and into $4\pi^0$ measured in $\pi^-\pi^+\pi^-\pi^+\pi^-$ [42,43] and into $\pi^-4\pi^0$ [44-47] pbar annihilations at rest in liquid $D_2$ are the dominant feature (well distinct from phase space) in the respective mass plots (see fig.5 left plate from ref. [43] and fig. 5 right plate from ref.[44-46]). The $4\pi^0$ data are particularly interesting because they do not suffer from any combinatorial background and there cannot be a $\rho\rho$ decay. These data show that $f_0(1370)$ decays dominantly into $\sigma\sigma$ (with a frequency twice that of the $\rho\rho$ decays and 6 times larger than ππ decays) and that 4 pion decays of $f_0(1370)$ are 10 times more frequent than 4 pion decays of $f_0(1500)$ [13].

The scenario which emerges is compatible with the expectation for a glueball to decay with equal rates into ππ and K$\bar{\text{K}}$ deriving from the equal couplings of gluons to n$\bar{\text{n}}$ and s$\bar{\text{s}}$ quarks [1,8] and expectations from QCD sum rules and low energy theorems for a scalar glueball to decay dominantly into σσ [48]. There is however the caveat that in the analysis of WA102 pp CEP data at 29 GeV the ρρ decay branching ratio of $f_0(1370)$ is found to be more than 4 times larger than the σσ decay branching ratio [49] and the ratios of the decay branching ratios of $f_0(1370)$ into ππ, K$\bar{\text{K}}$ and ηη to the decay branching ratios of $f_0(1370)$ into ππ are of the order of 1, 0,5 and 0,2 [50,51].

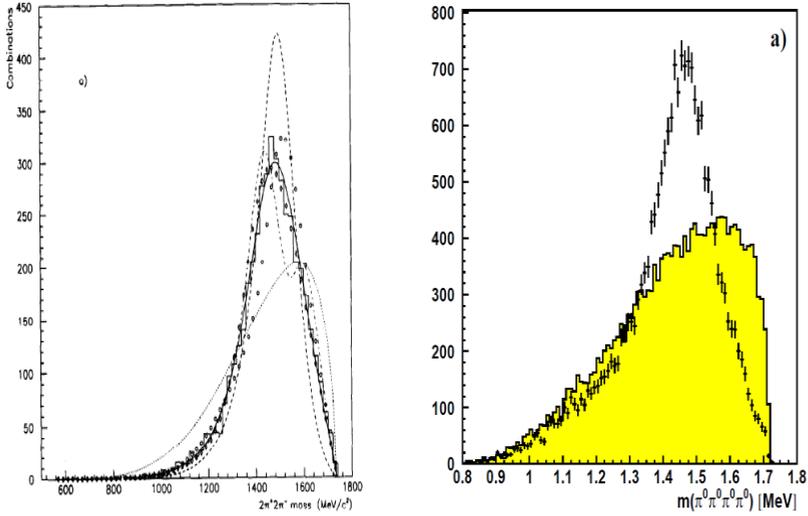

**Fig. 5.** $\pi^+\pi^-\pi^+\pi^-$ invariant mass in n$\bar{\text{p}}$ → $\pi^-\pi^+\pi^-\pi^+\pi^-$ annihilations in liquid $D_2$; the histogram shows data, the dotted line shows phase space (left plate, from ref.[43]). $4\pi^0$ invariant mass in n$\bar{\text{p}}$ → $\pi^-$ $4\pi^0$ annihilations at rest in liquid $D_2$; the peak –data points with errors- is mainly due to $f_0(1370)$ decays to $4\pi^0$, the colored histogram shows phase space (right plate, from ref. [44-46]).

## 4 Conclusions and prospects

The STAR data confirm indications of earlier CEP experiment at lower energies that moving to higher energies and lower ltl windows singes out pomeron-pomeron interactions, suppresses $2^{++}$ production and therefore produces spectra where all low mass $0^{++}$ mesons can appear with little background.

The clear STAR signal of the $f_0(1370)$ decays into $\pi^+\pi^-$ pairs confirms experimentally the data of p̄ annihilations at rest and shows that the σ meson and $f_0(1370)$ are distinct separate objects. The limited width of $f_0(1370)$ and its mass centered around 1370 MeV validate the analysis of data of p̄ annihilations at rest into 3 pseudoscalars that used the hypothesis of the existence of $f_0(1370)$ as an individual object [32,33,43-47,52,53]. The ratios of the decay branching ratios of $f_0(1370)$ into σσ, ρρ, ππ, KKbar and ηη to the decay branching ratios of $f_0(1370)$ into ππ are then of the order of 6, 3, 1, 1 and 0.02. There is however conflict between these decay branching ratios and the values obtained by WA102 in CEP measurements at 29 GeV [49-51]. The assessment of the values of the branching ratios of decays of the $f_0(1370)$ scalar meson into σσ and ρρ pairs and into ππ, K$\bar{\text{K}}$, ηη pairs of pseudoscalar mesons derived from p̄ annihilations at rest is discussed in more detail in ref. [54]. The decay properties of $f_0(1370)$ measured in pbar annihilations at rest and its production properties match the characteristics expected from an object that has a large gluon content [8-10,48].

CEP experiments at LHC at low ltl look extremely promising. The very large cms energy selects dominant pomeron-pomeron production. Low ltl select $0^{++}$ production within pomeron-pomeron production. Experiments equipped with precision detectors inside roman pots can approach the circulating beams and measure events with low ltl at both proton vertices. This is the case of ALFA-ATLAS and CMS-TOTEM, that have taken CEP data in 2015 with $\beta^*$ = 90 m optics of LHC and a ltl coverage which could go down to 0.03 GeV$^2$ [55]. With CEP data at LHC it should be possible in the medium term to establish with confidence the relative decay branching ratios of $f_0(1370)$ and $f_0(1500)$ (and also $f_0(1710)$) into $\pi^+\pi^-$, $K^+K^-$, $K_sK_s$ and compare with the values measured in antiproton annihilations at rest. Other important tasks will be to study of the narrow signal at 1450 MeV observed in CEP experiments at the Omega spectrometer in the $\pi^+\pi^-\pi^+\pi^-$ decay channel [56-59] and interpreted as possibly due to the interfering amplitudes of $f_0(1370)$ and $f_0(1500)$[58,59], to study the other 4 pion decay channels, and to compare the $\sigma\sigma$ and $\rho\rho$ decays of $f_0(1370)$ and $f_0(1500)$. In the longer term, measurements with larger $\beta^*$ and central detectors tuned to measure both low energy charged and neutral prongs the ltl window could be extended down to 0.003 GeV$^2$ and CEP could be measured in kinematical regimes where $0^{++}$ production should be definitively overwhelming and permit to do a complete study of all low energy scalars.